\documentstyle{article}

\def\ba{\begin{array}}
\def\ea{\end{array}}
\def\be{\begin{equation}}
\def\ee{\end{equation}}

\def\a{{\alpha}}
\def\s{{\sigma}}

\def\o{{\omega}}
\def\O{{\Omega}}
\def\bea{\begin{eqnarray}}
\def\eea{\end{eqnarray}}
\def\b{\beta}

\def\ti{\tilde}
\def\da{\dag}
\def\e{\eta}
\def\lm{\lambda}
\def\Lm{\Lambda}

\def\b{\beta}
\def\p{\emptyset}

\def\t{\tau}

\def\G{\Gamma}

\def\rd{{\rm d}}
\def\I{{\rm I}}
\def\sgn{{\rm sgn}}
\begin{document}
\begin{titlepage}
\hfill{}
\vskip 5 mm
\noindent{ \Large \bf
         Similarity transformation in one--dimensional reaction--diffusion
         systems; voting model as an example}

\vskip 1 cm
\noindent{Amir Aghamohammadi$^{1,3,a}$, Mohammad Khorrami$^{2,3,b}$}
\vskip 5 mm
{\it
  \noindent{ $^1$ Department of Physics, Alzahra University,
             Tehran 19834, Iran. }

  \noindent{ $^2$ Institute for Advanced Studies in Basic Sciences,
             P.O.Box 159, Gava Zang, Zanjan 45195, Iran. }

  \noindent{ $^3$ Institute for Studies in Theoretical Physics and
            Mathematics, P.O.Box  5531, Tehran 19395, Iran. }

  \noindent{ $^a$ mohamadi@theory.ipm.ac.ir}

  \noindent{ $^b$ mamwad@iasbs.ac.ir}
  }
\vskip 1 cm



\vskip 1cm

\begin{abstract}
The exact solution for a system with two--particle annihilation and
decoagulation has been studied. The spectrum of the Hamiltonian of the
system is found. It is shown that the steady state is two--fold degenerate.
The average number density in each cite $\langle  n_i(t)\rangle $ and the
equal time two--point functions  $\langle  n_i(t) n_j(t)\rangle $ are
calculated. Any equal time correlation functions at large times,
 $\langle  n_i({\infty}) n_j({\infty}) \cdots \rangle $,
is also calculated. The relaxation behaviour of the system toward its final
state is investigated and it is shown that generally it is exponential,
as it is expected. For the special symmetric case, the relaxation behaviour
of the system is a power law. For the asymmetric case, it is shown
that the profile of deviation from the final values is an
exponential function of the position.

\end{abstract}
\vskip 10 mm
\end{titlepage}
{\section {Introduction}}
In recent years, reaction--diffusion systems have been studied by many
people, using different methods. Among them are the field theoretic
methods, which allow for perturbative approaches to build up
correlations in low dimensions \cite{L,C}.
As mean field techniques can not be used for low dimensional
systems, people are motivated to study stochastic models in low
dimensions, which can be solved exactly.
Moreover, solving one dimensional systems should in principle
be easier. Applying a similarity transformation on an integrable model, one
may construct stochastic models, their integrability may be not obvious.
Recently, Some people have studied such transformations [3--8].

Exact results for some models in a one--dimensional lattice have
been obtained, for example in \cite{ADHR,AL}. In These cases, the time
evolution of the system is determined by a master equation \cite{KS}.
Models with no diffusion received less attention in the literature
[11--14]: It is said that unless the system has long--range reactions
\cite{SKB1,SKB2}, the time dependence involves exponential relaxation
rather than power law behaviour typical of the fast diffusion reactions.

In \cite{Sc}, a 10--parameter family of stochastic models has been studied.
In these models, the $k$--point equal time correlation functions
$\langle  n_i n_j\cdots n_k\rangle $ satisfy linear differential equations
involving no higher--order correlations.
These linear equations for the average density $\langle  n_i\rangle $ has
been solved. But, these set of equations may not be solved easily for
higher order correlation functions.
The spectrum is also partially obtained. The model which we address
in this article is a special case of that 10--parameter stochastic model.

In this work, we report the exact solution for a system with two--particle
annihilation and decoagulation. This model may be considered as a biased
voting model, in the sense that there are two different opinions. If the
two persons on two adjacent sites have different opinions, they may
interact so that their opinions become the same. The bias parameter
corresponds to the dominance of the left (or right) sight. In the absence
of bias, this system is equivalent to the zero--temperature Glauber model
\cite{Gl,S15}. This system is related to free fermion system, through a
similarity transformation, and hence is solvable. Note that the system
itself is not a free fermion system and can not be solved by applying only
Jordan--Wigner transformation.

When there is right--left symmetry, the average density decays to its final
value in the form of power law ($t^{-{1\over 2}}$). But in the general
case (biased model) it decays in the form of an exponential. Moreover,
the profile of the deviation of the average density from its final value
is not uniform but exponential in terms of the site number. In fact,
the parameter representing the right--left asymmetry, in some sense,
determines the dominance of the right sites over the left sites, or
vice versa.

The spectrum of the Hamiltonian of the system
is found. It is shown that the steady state is two--fold degenerate. The
probability of finding the system in each of these two states is determined
by the initial average density, and is time--independent. It is shown that
at large times, any $n$--point function is equal to the 1--point function,
which is position--independent.
\be
\langle  n_i(\infty )n_j(\infty )\cdots n_k(\infty )\rangle =\langle
n_i(\infty )\rangle =
{1\over L}\sum_m\langle  n_m(0)\rangle
\ee
This is due to the fact that the system has two steady states; either
completely full, or completely empty, as it will be shown. This means that
the mean--field approach does not work and this system is highly
correlated.

The scheme of the paper is as follows. In section 2, similarity
transformations relating stochastic systems to other (stochastic or
non--stochastic) systems are investigated. In section 3, a solvable model
is obtained through a similarity transformation on a free--fermion system.
The spectrum of the system is also obtained in this section. In section 4,
the 1--point function is calculated and its large--time behavior is
investigated. In section 5, the two--point function and its limiting
behavior is obtained. In section 6, The null vectors of the Hamiltonian are
obtained and from that the steady state of the system is obtained in terms
of its one--point function at $t=0$. Finally, in section 7 we consider the
next--to--leading term of the one--point function at large times, and from
this obtain the way the system relaxes to its final state.

{\section{Similarity transformations as a method for obtaining solvable
stochastic models}}
\noindent Here some standard material [2,3,5] is introduced, just to fix
notation. The master equation for $P(\s ,t)$ is
\be
{\partial\over\partial t} P(\s ,t)= \sum_{\t \ne \s}\big[ \o (\t\to \s )
P(\t ,t)-\o (\s \to \t )P(\s ,t) \big] ,
\ee
where $ \o (\t\to \s )$ is the transition rate from the configuration
$\t$ to $ \s$. Introducing the state
vector
\be
\vert P(t)\rangle  = \sum_{\s} P(\s ,t ) \vert \s \rangle ,
\ee
where the summation runs over all possible states of the system, one can
write the above equation in the form
\be \label{sh}
{\partial \over \partial t} \vert P\rangle ={\cal H} \vert P\rangle ,
\ee
where the matrix elements of ${\cal H}$ are
\bea
\langle   \s \vert {\cal H}\vert \t \rangle = \o ( \t \to \s ),
 \qquad \t \ne \s,\cr
\langle   \s \vert {\cal H}\vert \s \rangle = -\sum_{\t\ne \s}\o
( \s \to \t ).
\eea
The basis $\{\langle  \s \vert\}$ is dual to $\{ \vert\s \rangle \}$,
that is
\be
\langle   \s \vert  \t \rangle =\delta_{\s , \t}.
\ee
The operator is ${\cal H}$ is called a Hamiltonian, and it is not
necessarily hermitian. But, it has some properties.
Conservation of probability,
\be
\sum_{\s} P(\s ,t)=1,
\ee
shows that
\be
\langle  S \vert {\cal H}=0,
\ee
where
\be
\langle   S \vert =\sum_{\b} \langle   \b\vert .
\ee
So, the sum of each column of ${\cal H}$, as a matrix,  should be zero.
As $\langle S \vert$ is a left eigenvector of ${\cal H}$ with zero
eigenvalue,
${\cal H}$ has at least one right eigenvector with zero eigenvalue. This
state corresponds to the steady state distribution of the system and it
does
not evolve in time. If the zero eigenvalue is degenerate, the steady state
is not unique. The transition rates are non--negative, so the off--diagonal
elements of the matrix ${\cal H}$  are non--negative. Therefore, if a matrix
${\cal H}$ has the following properties,
\be\ba{l}
\langle   S \vert {\cal H}=0,\cr
\langle   \s \vert {\cal H}\vert \t \rangle \geq 0,
\ea \ee
then it can be considered as the generator of a stochastic
process. The real part of the eigenvalues of any matrix
with the above conditions should be less than or equal zero.

The dynamics of the state vectors (\ref{sh}) is given by
\be
\vert P(t)\rangle = \exp (t{\cal H}) \vert P(0)\rangle ,
\ee
and the expectation value of an observable ${\cal O}$ is
\be
\langle   {\cal O}\rangle (t)=\sum_{\s } {\cal O}(\s ) P(\s ,t)=\langle
S \vert {\cal O}
\exp (t{\cal H}) \vert P(0)\rangle .
\ee
If ${\cal H}$ is integrable, one can solve the problem, that is, one can
calculate the expectation values. Suppose now, that a Hamiltonian
is integrable but is not stochastic. There arises a question, whether or
not there exist exist a similarity transformation which transforms it to a
stochastic integrable Hamiltonian. Consider an integrable Hamiltonian
$\tilde {\cal H}$. The  similarity transformation
\be \label{bhb}
{\cal H}:={\cal B}\tilde {\cal H} {\cal B}^{-1}
\ee
leaves its eigenvalues invariant. Consider a special case:
The system
consists of a one dimensional lattice, with  nearest--neighbor interaction,
\be\label{47}
{\tilde {\cal H}}=\sum_{i=1}^L \ti{\cal H}_{i,i+1}.
\ee
Suppose, also, that the system is translation--invariant:
\be
\ti {\cal H}_{i\; i+1}=\underbrace{1\otimes\cdots\otimes 1}_{i-1}\otimes
{\ti H}\otimes\underbrace{ 1\otimes \cdots\otimes 1}_{L-i-1},
\ee
and we are using periodic boundary conditions. A simple class of
similarity transformations is then
\be \label{BB}
{\cal B}= \G_1\otimes \G_2 \otimes \cdots \otimes \G_L.
\ee
The simplest case is when all $\G_i$'s are the same.
In this case, if one can find $\G$ such that
\be
H=\G \otimes \G {\ti H}\G^{-1} \otimes \G^{-1}
\ee
is stochastic, then ${\cal H}$ defined through (\ref{bhb}) would be
stochastic.
A more general class of similarity transformations is obtained through
\be
\G_i:=\G (g)^i,
\ee
where $g$ should have the property
\be
[g \otimes  g , {\ti H}]=0.
\ee
In this case, one obtains
\be
H=(\G \otimes \G \ g)\  \ti H (\G \otimes \G \ g)^{-1}.
\ee
Define $\langle   s\vert $ to be the sum of all bra--states corresponding
to a single site. We then have
\be
\langle   S\vert =\underbrace{\langle   s\vert \otimes \cdots \otimes
\langle  s\vert}_{L}.
\ee
For $H$ to be stochastic, its off--diagonal elements should be
non--negative, and we must have
\be
\langle   s\vert \otimes \langle   s\vert H=0.
\ee

This shows that
\be
\langle   \a\vert \otimes \langle   \b\vert :=\langle   s\vert \G\otimes
\langle s \vert \G\ g,
\ee
should be an eigenvector with zero eigenvalue of $\ti H$, that is, $\ti H$
should have a decomposable left eigenvector. So, in order that this
prescription of constructing integrable stochastic model works, one must
begin with a Hamiltonian $\ti H$, the left eigenvector with zero eigenvalue
of which is decomposable. The real part of all other eigenvalues of $\ti H$
should, of course, be non--positive.

{\section{A one--parameter solvable system on the basis of a free--fermion
system}}
\noindent Consider the Hamiltonian
\bea \label{ffh}
\ti{\cal H}&=&\sum_{i=1}^L \{ {1+\e \over 2}[s_{i+1}^+s_i^--n_i(1-n_{i+1})]
           \cr
            &&+      {1-\e \over 2}[s_{i+1}^-s_i^+-n_{i+1}(1-n_i)]\cr
            & &     +\lm [s_{i+1}^-s_i^--n_in_{i+1}]\},
\eea
where $s^+, s^-,\  {\rm  and}\ n$ are
\be
s^+:=\pmatrix{0&1\cr 0&0},\qquad s^-:=\pmatrix{0&0\cr 1&0},\qquad
n:=\pmatrix{1&0\cr 0&0},
\ee
and the subscript $i$ represents the site, on which the operator acts.
This Hamiltonian describes the following processes
\bea
A\p \to \p A &&\qquad \hbox{with the rate} {1+\e\over 2}\cr
\p A\to A \p &&\qquad \hbox{with the rate} {1-\e\over 2}\cr
A A \to \p \p &&\qquad \hbox{with the rate} \lm .
\eea

This model has been recently studied. In the case $\lm =0$, the above model
describes an asymmetric exclusion process.
For $\lm =1$, the Hamiltonian is bilinear in terms of creation $s^+$ and
annihilation $s^-$ operators. This problem has been solved via a
Jordan--Wigner transformation \cite{S95,SSS}.
In the notation of the previous section the matrix form of
$\ti H$ is
\be
\ti H:=\pmatrix{-\lm &0&0&0\cr 0&-{1+\e\over 2}&{1-\e\over 2}&0\cr
 0&{1+\e\over 2}&-{1-\e\over 2}&0\cr \lm &0&0&0}
\ee
This matrix has two eigenvalues, 0 and -1, both of them are two--folded
degenerate. One of the zero left eigenvectors can be decomposed into a
tensor product. Doing the above mentioned procedure, this Hamiltonian
can  be transformed to another stochastic one. For this case, One can show
that the matrix
$g$ is the identity matrix, and the similarity transformation for all sites
become the same. This has been done in \cite{KPWH}.

One of the  left eigenvectors corresponding to the eigenvalue -1 has also
the desired property.
To use the prescription described in the previous section to construct
a stochastic Hamiltonian, we define  a new Hamiltonian,
\be
\ti H':=-\ti H -1,
\ee
and apply the similarity transformation on this new Hamiltonian. One of
the zero left eigenvectors of $\ti H'$ is
\be
(1\quad 0 \quad 0 \quad 0)=(1  \quad 0)\otimes (1\quad 0).
\ee
The similarity transformation should map $\langle   s \vert \otimes
\langle   s \vert $
to this vector:
\be
(1  \quad 1)\G\otimes (1  \quad 1)\G\ g=\a (1  \quad 0)\otimes (1  \quad 0)
\ee
So,
\be \ba{l}
(1  \quad 1)\G=\a \nu(1  \quad 0)\cr
(1  \quad 1)\G\ g={\a \over \nu } (1  \quad 0).
\ea \ee
Scaling $\G$ and $g$ does not alter the Hamiltonian $H$. So we can remove
$\a$ and $\nu $ by scaling the matrices $\G$ and $g$. Then the above
relation  gives some constraints on the elements of $\G$ and $g$. the
condition of positivity of rates, fixes $g$ and $\G$:
\be
\G={1\over 2}\pmatrix{1&-1\cr 1&1}
\ee
\be
g=\pmatrix{1&0\cr 0&-1}
\ee
The two site Hamiltonian, then, takes the following form
\be
H=\pmatrix{0&{1-\e\over 2}&{1+\e\over 2}&0\cr 0&-1&0&0\cr
 0&0&-1&0\cr 0&{1+\e\over 2}&{1-\e\over 2}&0}
 \ee
and ${\cal H}$ is
\bea  \label{ham}
{\cal H}&=&\sum_{i=1}^L \{ {1-\e \over 2}\left[n_is_{i+1}^++(1-n_i)
s_{i+1}^-)
\right]\cr && +{1+\e \over 2}\left[ s_{i}^+n_{i+1}+s_{i}^-(1-n_{i+1})
\right]\cr && -\left[ n_i(1-n_{i+1})-(1- n_i) \ n_{i+1}\right] \}.
\eea
This Hamiltonian describes the following processes
\bea
&A\p \to A A &\qquad {1-\e\over 2}\cr
&A\p \to \p \p &\qquad {1+\e\over 2}\cr
&\p A\to A A &\qquad {1+\e\over 2}\cr
&\p A \to \p \p &\qquad {1-\e\over 2} .
\eea
The Hamiltonian (\ref{ham})  is not quadratic in $s^+$ and $s^-$.
So, one can not map this Hamiltonian to a free fermion system, using a
Jordan--Wigner transformation. But the Hamiltonaian $\ti {\cal H}$ is
integrable and can be mapped to a free fermion system by a
Jordan--Wigner transformation. Consider the following Jordan--Wigner
transformation \cite{S95,SSS}
\bea
a_j&:=&Q_{j-1} s_j^- \cr
a_j^{\dag}&:=&Q_{j-1} s_j^+ \cr
Q_j&:=&\prod_{i=1}^j(-s_i^3).
\eea
It can be easily shown that the number operator at each site $n_i$ is,
in terms of new generators,
\be
n_i:={1+s_i^3\over 2}=a_i^{\dag}a_i
\ee
Using this transformation, The Hamiltonian $\ti {\cal H}$ takes
the following form
\be  \label{jwh}
\ti{\cal H}= \sum _{i=1}^L \left[{1-\e \over 2}a_i^{\da }a_{i+1}+{1+\e
             \over 2}a_{i+1}^{\da}a_i+a_{i+1}a_i-a_i^{\da}a_i \right],
\ee
$a_i$ and $a_i^{\da}$ fulfill the fermionic anti--commutation
relations
\be \label{fcr}  \ba{l}
\{ a_i,a_j\}=\{ a_i^{\da},a_j^{\da}\}=0\cr
\{ a_i,a_j^{\da}\}=\delta_{ij}.
\ea \ee
Note that it is in the limit $L\to \infty $ that the Jordan--Wigner
transformation we are using, works. Otherwise, there are some boundary
terms in (\ref{jwh}) as well. So, all the results we obtain hereafter, are
valid only in this limit. Now, introducing the Fourier transformation
\bea  \label{four}
a_j&:=&{1\over \sqrt{L}}\sum_k b_k \exp \{ {2\pi i\ j\ k \over L}\}\cr
a_j^{\da}&:=&{1\over \sqrt{L}}\sum_k b_k^{\da} \exp \{ {-2\pi i\ j\ k \over
L}\},
\eea
and substituting it in (\ref{fcr}), it is seen that
\bea \label{anti}
\{ b_k,b_l\}=\{ b_k^{\da},b_l^{\da}\}=0\cr
\{ b_k,b_l^{\da}\}=\delta_{kl}.
\eea
As a result, the Hamiltonian $\ti {\cal H}$ takes the form
\bea
\ti{\cal H}&=& \sum _{k}\left[ {1-\e \over 2}\exp ({2\pi i\ k \over L})
             +{1+\e \over 2}\exp ({-2\pi i\  k \over L})\right]b_k^{\da}b_k
             +b_{-k}b_k\exp ({-2\pi i\ k \over L})-b_k^{\da}b_k\cr
            &=&\sum _{k} \left[ \epsilon_k b_k^{\da}b_k
             -\ i \sin ( {2\pi  k \over L})b_{-k} b_k\right]
\eea
where
\be
\epsilon_k:=-1+\cos ({2\pi k \over L})-i\ \e \sin ({2\pi k \over L})
\ee
One can now, easily obtain the time dependence of $b_k$ and $b_k^{\da}$,
using (\ref{anti}) and ${\displaystyle{{\rm d}O \over {\rm d}t}}=[O,H]$
\bea   \label{bt}
b_k(t)=b_k(0) e^{\epsilon_k t}\cr
b_k^{\da}(t)=e^{-\epsilon_k t}\{ b_k^{\da}(0) -i \cot ({\pi k \over L})
[e^{(\epsilon_{k}+\epsilon_{-k}) t}-1] b_{-k}(0)\}
\eea
Now we return to our problem: determining the expectation values of a system
evolving with the Hamiltonian $\cal H$.
The expectation value of a quantity ${\cal O}$ is
\be\ba{ll}
\langle   {\cal O}\rangle (t)&=\langle   S \vert {\cal O}\exp (t{\cal H})
\vert P(0)\rangle \cr
       &=\langle   S \vert \exp (-t{\cal H}){\cal O}\exp ({\cal H}t) \vert
       P(0)\rangle .
\ea\ee
Substituting ${\cal H}=-{\cal B} \ti {\cal H} {\cal B}^{-1} -L{\bf 1}$,
where ${\bf 1}$ stands for the identity matrix, yields
\be
\langle   {\cal O}\rangle (t)=\langle   \O \vert \ti {\cal O}(-\ti t)
{\cal B}^{-1}\vert P(0)\rangle
\ee
where
\be
\ti {\cal O}:= {\cal B}^{-1} {\cal O} {\cal B},
\ee
\be
\ti{\cal O}(-\ti t):= e^{t{\ti {\cal H}}}{\cal O}e^{-t{\ti {\cal H}}}
\ee
and
\be
\langle   \O \vert :=\pmatrix{1&0} \otimes \pmatrix{1&0}\otimes \cdots
\pmatrix{1&0}.
\ee
The main expectation values of interest are the correlation functions
of $n_i$`s. To determine these, we use
\be \ba{l}
\G^{-1} n \G={1\over 2} ( 1-s^+ -s^-)\cr
({\G g})^{-1} n {\G g}={1\over 2} ( 1+s^+ +s^-)
\ea \ee
So
\be
{\cal B}^{-1} n_i {\cal B} ={1\over 2} [ 1-(-1)^i(s^+ +s^-)].
\ee
Now, we want to calculate the expectation value of ${\cal O}$ where
\be
{\cal O}:=n_{i_m}\cdots n_{i_2}n_{i_1}, \qquad i_1\langle   i_2\langle
\cdots \langle   i_m.
\ee
We have
\be
\ti{\cal O}={1\over 2^m}\big[ 1-(-1)^{i_m}(s_{i_m}^+ +s_{i_m}^-)\big]\cdots
             \big[ 1-(-1)^{i_1}(s_{i_1}^+ +s_{i_1}^-)\big].
\ee
Using the Jordan--Wigner transformation, one arrives at
\be
\ti{\cal O}={1\over 2^m}\big[ 1-(-1)^{i_m}Q_{i_m-1}(a_{i_m}^{\da}+
a_{i_m})\big]\cdots \big[ 1-(-1)^{i_1}Q_{i_1-1}(a_{i_1}^{\da}+a_{i_1})\big].
\ee
It is easy to check that $\langle   \O\vert Q_i= (-1)^i\langle   \O\vert $.
So in calculating
$\langle   {\cal O}\rangle $, one can use ${\cal O'}$ instead of
$\ti{\cal O}$:
\be
{\cal O'}:={1\over 2^m}\big[ 1+(a_{i_m}^{\da} +a_{i_m})\big]
           \cdots \left[ 1+(a_{i_1}^{\da}+a_{i_1})\right].
\ee
Instead of ${\cal O'}$, It is enough to set ${\cal O''}$ in the expectation
value of ${\cal O}$, where
\be \label{o''}
{\cal O''}:={1\over 2^m}\big( 1+a_{i_m}^{\da}\big)
           \cdots \big( 1+a_{i_1}^{\da}\big).
\ee
To prove this, one should use $\langle   \O\vert \ti {\cal H}=-L\langle
\O \vert $ and $\langle   \O\vert a_i(0)=0$.

{\section{The one--point function}}
\noindent As the first example, consider the one--point function $\langle
 n_m(t)\rangle $:
\be\label{an}
\langle   n_m(t)\rangle ={1\over 2}\langle   \O\vert [1+a_m^{\da}(-\ti t)]
{\cal B}^{-1}\vert P(0)\rangle .
\ee
Using the Fourier transformation (\ref{four}), the time dependence of
$b_k^{\da}$ (\ref{bt}), and remembering  $\langle   \O\vert b_k(0)=0$,
we obtain
\be
\langle   n_m(t)\rangle ={1\over 2}+{1\over 2\sqrt{L}}\sum_k
e^{-2\pi i km\over L}
\langle   \O\vert b_k^{\da}(0){\cal B}^{-1}\vert P(0)\rangle
e^{\epsilon_k t}
\ee
Now, we use the inverse Fourier--, and Jordan--Wigner--transformations,
and arrive at
\be
\langle   n_m(t)\rangle ={1\over 2}+{1\over 2L}\sum_{k,j} e^{2\pi i k(j-m)
\over L}
\langle   S\vert{\cal B}s_j^+{\cal B}^{-1}\vert P(0)
\rangle e^{\epsilon_k t}(-1)^{j-1}.
\ee
This  can be written in a simpler form, using
\be
{\cal B}s_j^+ {\cal B}^{-1}=(-1)^{j-1}{2n_j-1+s_j^- -s_j^+\over 2}
\ee
and
\be
\langle   s\vert (2n -1)=\langle   s\vert (s^- -s^+).
\ee
One then arrives at
\be
\langle   n_m(t)\rangle ={1\over 2}+{1\over 2L}\sum_{k,j}
 e^{2\pi i k(j-m)\over L}
\langle   S\vert (2n_j(0) -1)\vert P(0)\rangle e^{\epsilon_k t}.
\ee
Using \ $ \langle   S\vert P(0) \rangle =\sum_{\s}P(\s ,0)=1$, \
 one arrives at
\be
\langle   n_m(t)\rangle =\sum_{j}\Lm_{mj}(t) \langle   n_j(0)\rangle ,
\ee
where
\be
\Lm_{mj}(t):={1\over L}\sum_{k}e^{2\pi i k(j-m)\over L}e^{\epsilon_k
t}.
\ee
Now, consider the limit $t\to \infty$. In this limit, the only contribution
in the above summation comes from the term $k=0$. So,
\be \label{ani}
\lim _{t\to \infty} \langle   n_m(t)\rangle ={1\over L}\sum_{j}
\langle   n_j(0)\rangle ,
\ee
which shows that in the limit $t\to \infty$, the expectation value of the
number of particles in any site tends to the average of the initial value
of this quantity. In the last section, we will find the next leading term
of $\langle   n_i(t)\rangle $, for large times.

Now, we want to calculate the expectation value of the number of particles
in the site $j$ in the limit $L\to \infty$. First, we calculate $\Lm_{mj}$
in this limit. To do so, we define $z:=\exp ({i{2\pi k\over L}})$. We then
(in this limit) arrive at
\be
\Lm_{mj}(t)=e^{-t}\oint {{\rm d}z\over 2\pi i z} z^{j-m}\exp[t( {1-\e
\over 2}z+{1+\e \over 2}z^{-1})].
\ee
Changing the variable $z$ to $w :=z\sqrt{{1+\e \over 1-\e}}$, the
above integral takes the form
\be
\Lm_{mj}(t)=e^{-t}\left({1-\e \over 1+\e }\right)^{j-m\over 2}\oint
{{\rm d}w\over 2\pi i w} w^{j-m}\exp [{t\sqrt{1-\e^2\over
2}}(w+w^{-1})],
\ee
or, using the change of variable $w:=e^{i\theta}$,
\be
\Lm_{mj}(t)=e^{-t}\left({1-\e\over 1+\e}\right)^{m-j\over 2}\int_{0}^{2\pi}
{{\rm d}\theta\over 2\pi } e^{i(j-m)\theta +t\sqrt{1-\e^2}\cos \theta }.
\ee
The above integral is an integral representation of the modified Bessel
function:
\be
\Lm_{mj}(t)=\left({1-\e\over 1+\e}\right)^{m-j\over 2}{\rm I}_{m-j}
(t\sqrt{1-\e^2})e^{-t}.
\ee
$\langle   n_m(t)\rangle $, in the limit $L \to \infty$, is then
\be \label{007}
\langle   n_m(t)\rangle =\sum_{j}\left({1-\e\over 1+\e}\right)^{m-j\over 2}
{\rm I}_{m-j}
(t\sqrt{1-\e^2})e^{-t} \langle   n_j(0)\rangle .
\ee

{\section{The two--point function}}
\noindent The other quantity which we want to calculate is
$\langle   n_m(t)n_l(t)\rangle $.
Without loss of generality, one may assume $(m > l)$.
To calculate this, we use (\ref{o''}), which gives
\bea \label{nn}
&\langle   n_m(t)n_l(t)\rangle ={1\over 4}\langle   \O\vert
[1+a_m^{\da}(-\ti t)][1+a_l^{\da}(-\ti t)]
{\cal B}^{-1}\vert P(0)\rangle \cr
&\ \ \ \ \ \ \ \ =-{1\over 4}+{1\over 2}[\langle   n_m(t)\rangle +\langle
 n_l(t)\rangle ]+{1\over 4}\langle   \O\vert a_m^{\da}
(-\ti t) a_l^{\da}(-\ti t){\cal B}^{-1}\vert P(0)\rangle .
\eea
The main thing is to calculate the last term. To do this, we first use
the Fourier transformation of $a_i^{\da}$'s,
\be
\langle   \O\vert a_m^{\da}(-\ti t)a_l^{\da}(-\ti t)){\cal B}^{-1}
\vert P(0)\rangle =
{1\over L}\sum_{k,p} e^{-i\ 2\pi {(km+pl)\over L}}
\langle   \O\vert b_k^{\da}(-\ti t)b_p^{\da}(-\ti t){\cal B}^{-1}\vert
P(0)\rangle ,
\ee
and then substitute the time dependence of $b_k^{\da}$s.
\bea
\langle   \O\vert a_m^{\da}(-\ti t)a_l^{\da}(-\ti t)){\cal B}^{-1}\vert
P(0)\rangle &=&
{1\over L}\sum_{k,p} e^{-i\ 2\pi {(km+pl)\over L}+(\epsilon_k +
\epsilon_p )t}\langle  \O\vert b_k^{\da}(0)\cr &&\left[b_p^{\da}(0)+i \cot
({\pi p\over L})\left(1-e^{-(\epsilon_p +\epsilon_{-p} )t}\right)
b_{-p}(0)\right]{\cal B}^{-1}\vert
P(0)\rangle .
\eea
Now we use inverse Fourier transformation for the
$b_k^{\da}b_p^{\da}$ term. The other term is easily summed. We arrive at,
\bea
\langle   \O\vert a_m^{\da}(-\ti t)a_l^{\da}(-\ti t){\cal B}^{-1}\vert
P(0)\rangle &=&
\sum_{r,s} \Lm_{mr}(t)\Lm_{ls}(t)\langle   \O\vert a_r^{\da}
a_s^{\da}{\cal B}^{-1}\vert P(0)\rangle \cr &&+{i\over L}
\sum_{k} e^{i\ 2\pi {k(l-m)\over L}}\cot ({\pi k\over L})
\left(1-e^{(\epsilon_k +\epsilon_{-k} )t}\right),
\eea
or,
\bea
\langle   \O\vert a_m^{\da}(-\ti t)a_l^{\da}(-\ti t){\cal B}^{-1}\vert
P(0)\rangle &=&
\sum_{k,p,r,s} \Lm_{mr}(t)\Lm_{ls}(t)\langle    (2n_r-1)(2n_s-1)\rangle _0
 \sgn (r-s)\cr &&
+{i\over L}\sum_{k} e^{i\ 2\pi {k(l-m)\over L}}\cot ({\pi k\over L})
\left( 1-e^{(\epsilon_k +\epsilon_{-k} )t}\right) ,
\eea
where we have used the definition of $\Lm_{ij}$, and $\langle
 \cdots \rangle _0$
means the expectation value at the initial
time. Adding all terms in (\ref{nn}) together, one arrives at
\bea
\langle   n_m(t)n_l(t)\rangle &=&-{1\over 4}(\langle   n_m(t)\rangle +
\langle   n_l(t)\rangle )+\sum_{r,s}\Lm_{mr}(t)
\Lm_{ls}(t) \langle   (n_r-{1\over 2})(n_s-{1\over 2})\rangle _0
\sgn (r-s)\cr &&
+{i\over 4L}\sum_{k} e^{i\ 2\pi {k(l-m)\over L}}\cot ({\pi k\over L})
\left(1-e^{(\epsilon_k +\epsilon_{-k} )t}\right).
\eea
The last term is independent of initial conditions, So we can
calculate it for a special case, e.g. $\vert P(0)\rangle =\vert 0\rangle $.
Then, the final result is
\be
\langle   n_m(t)n_l(t)\rangle ={1\over 2}(\langle   n_m(t)\rangle +\langle   n_l(t)\rangle )+\sum_{r,s}\Lm_{mr}(t)\Lm_{ls}(t)
\sgn (r-s) \langle   n_rn_s-{n_r+n_s\over 2}\rangle _0
\ee
For large times, it is seen that
\be
\lim _{t \to \infty}\langle   n_m(t)n_l(t)\rangle =\langle   n(\infty )
\rangle
\ee

{\section{Null vectors of the Hamiltonian, the steady state of the system,
and the $n$--point function}}
\noindent Now we want to study the null eigenvectors of the Hamiltonian
${\cal H}$.
It is easy to see that the Hamiltonian (\ref{ham}) has at least two null
eigenvectors, which means that the steady state is not unique. These
states are one in which all sites are occupied, and one in which  no
site is occupied. One can check this easily by acting the Hamiltonian
(\ref{ham}) on these states. It was shown that the Hamiltonian
$\tilde{\cal H}$ may be written as
\bea
\ti{\cal H}&=&\sum _{k} \left[ \epsilon_k b_k^{\da}b_k
             -\ i \sin ( {2\pi  k \over L})b_{-k} b_k\right]\cr
            &=&\sum _{k> 0} \left[ \epsilon_k b_k^{\da}b_k
           + \epsilon_{-k} b_{-k}^{\da}b_{-k}
             -2\ i \sin ( {2\pi  k \over L})b_{-k} b_k\right]
            +\epsilon_0 b_0^{\da}b_0.
\eea
This Hamiltonian is obviously block diagonal. In each four dimensional
block,
one can choose a basis $\{ \vert 0\rangle ,b^{\da}_k \vert 0\rangle ,
 b^{\da}_{-k}\vert 0\rangle ,
b^{\da}_k b^{\da}_{-k} \vert 0\rangle  \}$. The eigenvalues of this four
dimensional
block are $0,\ \epsilon_k,\ \epsilon_{-k},\ \epsilon_k+\epsilon_{-k}$, or
${\cal N}_k\epsilon_k +{\cal N}_{-k}\epsilon_{-k}$, where ${\cal N}$'s are
zero or one. The eigenvalues of the Hamiltonian are, therefore,
\be
\tilde E\{ {\cal N } \}=\sum_k{\cal N}_k\epsilon_k.
\ee
From this, one can obtain the eigenvalues of ${\cal H}$ as
\bea
E\{ {\cal N } \}&=&-\sum_k({\cal N}_k\epsilon_k+1)\cr
                &=&\sum_k(-{\cal N}_k+1)\epsilon_k.
\eea
Here we have used
\be
\sum_k(1+\epsilon_k)=0.
\ee
Now, it is easy to see that $E$ is zero iff $1-{\cal N}_k=0, \qquad
\forall k\ne 0$. This shows that the null eigenvector is two--fold
degenerate. As the final state is two--fold degenerate, and it is known
that the totally full state $(\vert \O \rangle :=\pmatrix{1\cr 0}\otimes
\pmatrix{1\cr 0}\otimes\cdots
\pmatrix{1\cr 0})$ and totally empty state
$(\vert 0 \rangle :=\pmatrix{0\cr 1}\otimes\pmatrix{0\cr 1}\otimes\cdots
\pmatrix{0\cr 1})$ are null eigenvectors of the system, we have
\be
\vert P(\infty )\rangle =\a \vert 0\rangle  +\b \vert \O \rangle ,
\ee
where $\a +\b =1$. Using (\ref{ani}), and
\bea
\langle   S\vert n_i \vert 0\rangle =0,\cr
\langle   S\vert n_i \vert \O\rangle =1,
\eea
it is seen that
\be
\b =\langle   n(\infty )\rangle ={1\over L}\sum_j \langle   n_j(0)
\rangle =:\rho_0
\ee
and
\be
\a =1-\rho_0.
\ee
From this, one obtains
\be
\vert P(\infty )\rangle =[1-\rho_0 ]\vert 0\rangle  +\rho_0 \vert \O
\rangle .
\ee
Using this, it is easy to find all $m$-point functions in the limit $t \to
\infty $. We have
\bea
\langle   n_{i_m}\cdots n_{i_1}\rangle (\infty )&=&\langle
 S\vert n_{i_m}\cdots n_{i_1}\vert
P(\infty )\rangle \cr
&=& \rho_0\cr
&=&\langle   n(\infty )\rangle \cr
&\ne& \langle   n(\infty )\rangle ^m
\eea
This clearly shows that the mean--field approximation does not work here.

{\section{Relaxation of the system toward its steady state}}
\noindent It was shown that in the limit $t\to\infty$, the expectation of
the number of particles in any site tends to the average of the initial
value of this quantity. Now, we want to study the behaviour of the system
at large times. Starting from (\ref{007}), and representing
$\langle   n_j(0)\rangle $ by its
Fourier transform, we have
\be
\langle   n_m(t)\rangle =e^{-t}\int_0^{2\pi}{{\rd u}\over{2\pi}}
\sum_j\left(
{{1-\eta}\over{1+\eta}}\right)\I_{m-j}(t\sqrt{1-\eta^2})e^{-iuj}[f(u)+2\pi
\bar n\delta (u)],
\ee
where $[f(u)+2\pi\bar n\delta (u)]$ is the Fourier transform of
$\langle   n_j(0)\rangle $,
and $\bar n$ is the average density. We have extracted this part of the
Fourier transform, so that the remaining is a smooth function of $u$.
Then, $f(u)$ denotes the Fourier transform of the deviation
$ \langle   n_j(0)\rangle - \bar n$. The summation on $j$ is easily done,
using
\be
\sum_n x^n \I_n(y)=e^{(y/2)(x+1/x)}.
\ee
So, one arrives at
\be
\langle   n_m(t)\rangle =\bar n+\int_0^{2\pi}{{\rd u}\over{2\pi}}
e^{t\epsilon(u)}f(u) e^{-imu},
\ee
where $\epsilon(u)$ is the same as $\epsilon_k$ with $k=Lu/(2\pi)$. The
above integral is simplified for large times, using the steepest descent
method. Using the change of variable $z:=e^{iu}$, the integral becomes
\be
\langle   n_m(t)\rangle =\bar n+\oint{{\rd z}\over{2\pi i z}}
e^{t[-2+(1-\eta)z+(1+\eta)
z^{-1}]/2}\tilde f(z)z^{-m},
\ee
where the integration contour is the unit circle. The multiplier of $t$ in
the exponent is stationary at
\be
z_1=\sqrt{{1+\eta}\over{1-\eta}},
\ee
and
\be
z_2=-\sqrt{{1+\eta}\over{1-\eta}}.
\ee
As the real part of this multiplier is larger at $z=z_1$, the integral gets
its main contribution from this point. (This point is not on the
integration contour. But, assuming $\tilde f$ to be analytic, one deforms
the integration contour so that it passes from $z_1$, and then uses the
steepest descent method.) We arrive at
\be
\langle   n_m(t)\rangle -\bar n\sim{1\over{\sqrt{t}}}\left({{1-\eta}
\over{1+\eta}}\right)
^{m/2}e^{t(\sqrt{1-\eta^2}-1)}.
\ee
The effect of the Fourier transform $\tilde f$, and the second derivative
of the multiplier of $t$ in the exponent is a multiplier independent of $m$
and $t$. Two general features, independent of the initial condition, are
seen from the above relation. First, the decay to the final state is not
in the form of a power law, but in the form of an exponential. It becomes a
power law only in the symmetric case $\eta =0$. Second, if $\eta >0$,
the expectation at the rightmost sites tends rapidly to its final value.
That is, the profile of the deviation from the final value is decreasing
with respect to $m$. This is so, since in this case the two--site
reaction is favorable to the state where the left site changes so that
it becomes identical to the right site. This means that cases where the
right--site changes is less probable than cases where the left site
changes. So, the right site arrives earlier to its final state.
This expression seems to be unbounded for either $m\to\infty$ or
$m\to -\infty$. For any fixed $t$, this is true. But it simply means that
in order that this term represents the leading term for some $m$, $t$ must
be greater than some $T$, which {\it does depend} on $m$.

\newpage

\end{document}